\documentstyle[preprint,aps]{revtex}
\begin{document}
\title{Optical properties of the vibrations in charged C$_{60}$
molecules}
\author{Arthur P. Smith$^a$ and George F. Bertsch$^b$}
\address{
$^a$
Chemistry Department, Box 351700, University of Washington, Seattle,
WA 98195-1700
}
\address{
$^b$
Physics Department and Institute for Nuclear Theory, Box 351560,
University of Washington, Seattle, WA 98195-1560
}
\maketitle

\begin{abstract}
The transition strengths for the four infrared-active vibrations of
charged C$_{60}$ molecules are evaluated in self-consistent
density functional theory using the local density
approximation. The oscillator strengths for the second and fourth
modes
are strongly enhanced relative to the neutral C$_{60}$ molecule, in
good agreement with the experimental observation of ``giant
resonances'' for those two modes. Previous theory, based on a
``charged phonon'' model, predicted a quadratic dependence of the
oscillator
strength on doping, but this is not borne out
in our calculations.
\pacs{PACS categories: 33.20.Ea, 36.40.Wa, 78.30.Hv, 36.40.Vz}
\end{abstract}

\section{Introduction}

Recent experimental measurements\cite{Fu,Martin,Pichler}
of infrared absorption in
A$_x$C$_{60}$ (A = K, Rb), ($x$ = 0, 1, 3, 4, 6) have demonstrated
the strong dependence on doping of the oscillator strengths for
several of the T$_{1u}$ infrared-active modes of the C$_{60}$
molecule.
The second and fourth modes are enhanced by roughly 100, while the
first and third show little or no change in oscillator strength
in going from undoped C$_{60}$ to C$_{60}^{6-}$.

The ``charged phonon'' model introduced by Rice and Choi
\cite{Rice+Choi}
obtains this increase in oscillator strength from the
stronger coupling of the vibrational mode to an electronic transition.
In particular the transition
of the dopant electrons between the $t_{1u}$ and the
higher-lying $t_{1g}$ molecular orbitals could enhance the dipole
associated with the vibrational modes. The result is, in addition
to a softening of the phonon modes proportional to the number of
dopant electrons $x$, an enhancement
of the oscillator strengths proportional to $x^2$.
Friedman examined the effect of doping in a tight-binding model
\cite{Friedman,F2}.  One parameter set did obtain the observed enhancement
in the fourth mode, but the second mode could not be explained.
Furthermore, experimental
evidence for the $x^2$ doping dependence appears inconclusive.

In a previous paper\cite{Bertsch} we examined the oscillator
strengths for the
transitions in undoped C$_{60}$, and found that the calculations
using a frozen-phonon approach in the local density approximation
gave good agreement with experiment on the magnitude of the oscillator
strengths, particularly when compared with a tight-binding approach.
In the present paper we apply the same self-consistent density
functional
technique to charged C$_{60}$ molecules, evaluating the dipole moment
induced on the system by the same vibrational distortion used
for the undoped system.

\section{Theory}
   As described, we make an adiabatic approximation to
calculate the oscillator strength associated with the vibrations.
We will calculate by distorting the C$_{60}$ structure according to
the atomic displacements in the vibrational modes, and
then evaluating the dipole moment of the structure.  Defining
the displacement of an atom $i$ in the mode $\alpha$ by the
vector $\vec{d}^\alpha_i$, the oscillator strength is given by
the formula
\begin{equation}
f_\alpha = { m_e (D_\alpha)^2 \over M_C \sum_i d^\alpha_i\cdot
d^\alpha_i }
\label{eq3}
\end{equation}
where $D_\alpha$ is the dipole moment in the distorted
structure and $m_e,M_C$ are the electron and carbon masses,
respectively.  The displacement vectors are the same as
we used for undoped C$_{60}$ \cite{Bertsch}, calculated from the
phenomenological model of C$_{60}$ vibrations in ref. \onlinecite{We88}.
Equation \ref{eq3} is of course the standard way to obtain
absorption strengths, although it is usually presented
in terms of infinitesimal displacements ({\it i.e.} the dipole
derivatives), rather than the finite displacements we use.

The electrons of the system are described by the Kohn-Sham
density functional approach\cite{Ko76}, in which an effective
one-electron
problem is introduced that depends on a self-consistent charge
density through an approximate ``exchange-correlation'' functional,
which takes into account the electron-electron interactions. We
use the local-density approximation (LDA)\cite{Pe81} for this
exchange-correlation functional, which has proven remarkably
successful in describing the charge distributions and energies of
a wide variety of solid and molecular materials.
Our calculation of the electron charge density for these fixed ionic
positions uses a version of the Car-Parrinello electronic
structure technique\cite{Ca85,Pa92} implemented in a code provided
by Wiggs\cite{Wi94}, recently modified to facilitate
rapid convergence to the ground state\cite{Gi95}.
The same pseudopotential\cite{Tr91} and plane-wave basis set
(with 35 Ry cutoff) was used here as in our earlier work. The
discretization of the plane wave basis set is equivalent to treating
the problem in a cubic box with periodic boundary conditions. The
equivalent box length in our calculations was 23 atomic units, or
12.2 \AA. Wave functions for the charged C$_{60}$ molecules were
calculated self-consistently for the excess charge and distorted
molecule conditions, with double occupancy assumed for all the
lowest-energy
orbitals, except that the highest occupied level contained only one
electron
for the odd-charge molecules. Note that the periodic boundary
conditions
imply that the Hamiltonian effectively has a uniform positive
background
charge, instead of the discrete positive ion charges that would
be present in the real M$_x$C$_{60}$ material.

This technique was then applied to calculate the charge density
and dipole moment for each of the frozen T$_{1u}$ modes for the C$_{60}$
ions, using a displacement of net amplitude of 1\AA, which in our
earlier work was shown to be still within the regime where the
moment varies linearly with the displacements. We
estimate the random errors (due to insufficient convergence or
box size)
in the resulting dipole moments to be roughly 0.1 e\AA, or between
6\% for the largest moments and 100\% for the smallest ({\it ie.}
the smallest $f$-strengths we calculate are not significantly
different from zero).

Before proceeding to the oscillator strengths, we note that the
vibrational frequencies are readily obtained from the energy differences
due to these displacements, and we have tabulated the results
for neutral and charge -6 C$_{60}$ in Table \ref{table1}. The
pronounced mode softening apparent in the experiments for the
first and fourth modes is absent in our calculated results - in
fact we find the first mode increases in frequency. This is most
likely a consequence of our assumption of identical vibrational
displacements in the neutral and charged systems.

The results for the oscillator strengths are much more conclusive.
As can be seen from Figure \ref{fig1}, none of the oscillator
strengths we obtain increases quadratically with $x$. In
agreement with experiment, the second and fourth modes are
strongly enhanced in strength. Our calculations find the first
and third to actually diminish in oscillator strength with
doping. The experiments do not clearly demonstrate this,
although the fact that these peaks are not observed in some
of the experiments at intermediate doping is evidence that they may
have decreased in amplitude.

Figure \ref{fig2} shows the relation between our calculations for
the neutral and charged C$_{60}$ molecules and the available
experiments on
the undoped and doped systems. We have converted the experimental
measurements to units corresponding to our theoretical isolated
C$_{60}$
$f$-strengths as outlined in the appendix. Theory and experiment
are generally in agreement to within a factor of 2, which is
remarkable
because the strengths vary over 2 orders of magnitude, and
different experiments themselves have discrepancies of
roughly the same size. However, there are a few modes
which were observed experimentally to have appreciable strength even
though the theoretical strengths were much smaller - in particular
the first mode for C$_{60}^{3-}$ and the first and third modes for
C$_{60}^{6-}$. This could be due to a breakdown in the theoretical
approximations, or possibly due to contamination of the experimental
samples by some fraction of neutral or charge -1 C$_{60}$.
The most significant disagreement appears to be for the fourth
mode of the
charge -6 species, where the theoretical strength is between 3
and 5 times
smaller than the experimental numbers. The most likely explanation of
this is the breakdown of our theoretical assumption that the atomic
motions
in the vibrational modes are independent of charge state. However
there
may be other approximations (such as ignoring the actual positions
of the
compensating charges, or the conversion from solid to molecular
strengths)
that are invalid for this particular data point.

\section{Conclusion}

Our calculations based on the Kohn-Sham density functional theory
thus appear to be in excellent agreement
with the available experiments. In particular, this self-consistent
approach correctly predicts the strong enhancement of the
second and fourth modes in the doped systems. The quantitative
agreement is generally within a factor of two, which is also roughly
the extent to which different experiments agree. However, all but
one of
the modes for the C$_{60}^{6-}$ ion are significantly weaker in the
theoretical calculations than in the experiments, possibly
indicating that
for such a highly charged ion the approximations we have made are
no longer
suitable. The experimental observation of the weak first and
third modes could also be ascribed to sample inhomogeneities and
resulting contamination from other charge states, however.

We note here that the net dipole moment for a particular oscillation
mode results from a very delicate cancellation of contributions from
almost all of the electronic states in the system - most of the states
have contributions to the total dipole that are roughly of the same
magnitude (one electron-\AA) as the final dipole moment. Thus the
charged-phonon model of Rice and Choi\cite{Rice+Choi} appears
to be oversimplified.  As shown in our Figure \ref{fig1}, the
quadratic doping dependence of the strengths predicted by that model
is not observed for any of the modes in the theoretical calculations.
Unfortunately, our calculations do not suggest any simple explanation
of the doping dependence - it appears that the only way to
obtain meaningful theoretical numbers is through such a
self-consistent calculation of the response of the electrons to
the vibrations.

\section{Acknowledgment}

We thank J. Wiggs for providing us the LDA program used for
these calculations.  This work was supported partly by
the Department of Energy under Grant FG06-90ER-40561 and
partly by the National Science Foundation under award number
CHE-9217294 (CARM).

\appendix
\section*{The experimental oscillator strengths}
In the published experimental work, the absolute strengths of the
absorption peaks are generally expressed in terms of an effective
plasma
frequency $\omega_p$ for each mode, {\it ie.} a contribution
\begin{equation}
4\pi \chi_j(\omega) = \omega_p^2/(\omega_j^2 - \omega^2 -
i \omega \gamma_j)
\end{equation}
associated with mode $j$ to the dielectric function of the solid
material,
where $\omega_j$ is the observed frequency of the resonance.
Theoretically,
the dielectric response $\epsilon(\omega)$ of the solid derives from
the polarizability $\gamma_{mol}(\omega)$ of the single molecule
through the Clausis-Mossotti relation:
\begin{equation}
\epsilon(\omega) = 1 + 4\pi {N \gamma_{mol}(\omega) \over
			1 - {4\pi \over 3} N \gamma_{mol}(\omega)}
\label{eq:eps-solid}
\end{equation}
where $N$ is the density of molecules in the solid. The fundamental
$f$-strengths we calculate are associated with the single-molecule
polarizability (ignoring damping) due to all electromagnetic resonance
modes in the molecule:
\begin{equation}
\gamma_{mol}(\omega) = {e^2 \over m} \sum_{j} {f_j \over
			(\omega_{j0}^2 - \omega^2)}
\end{equation}
The denominator in equation (\ref{eq:eps-solid}) shifts
the resonant frequency from $\omega_{j0}$ to $\omega_j$, and also
changes
the overall strength by a factor depending on the other modes.  These
other modes can be absorbed into a real net background contribution
to the
dielectric function that manifests itself simply as a value of the
refractive index $n$ different from 1
(recall $\epsilon = (n + ik)^2$).
The resulting relation between the effective plasma frequency
$\omega_p$
for a mode and the associated molecular $f$-strength $f_j$ is:
\begin{equation}
f_j = {9 \over (n^2 + 2)^2} {m \over 4\pi N e^2} \omega_p^2.
\label{eq:f-mol}
\end{equation}

We have used equation (\ref{eq:f-mol}) directly to convert the
results of
reference \onlinecite{Pichler}. The experimental results of reference
\onlinecite{Martin} requirement multiplication of the tabulated $S$
by the vibrational frequency $\omega_0^2$ to obtain $\omega_p^2$, and
then $f_j$. Reference \onlinecite{Chase} was
treated in our previous paper\cite{Bertsch}. Reference \onlinecite{Fu}
only provided relative strengths between the charge -6 and neutral
species: we assumed the numbers for the neutral species were
identical to that of reference \onlinecite{Martin}, and plotted
the resulting $f$-strengths for the charged molecule.

\begin{table}
\caption{Energies of the vibrations, in eV.}
\begin{tabular}{ccccc}
 &\multicolumn{2}{c}{C$_{60}$} & \multicolumn{2}{c}{C$_{60}^{6-}$} \\
 \cline{2-3} \cline{4-5}
Label&Experiment & LDA & Experiment & LDA \\
\tableline
$T_{1u}(1)$& 0.065 & 0.065 &  0.058 & 0.069 \\
$T_{1u}(2)$ & 0.071 & 0.070 & 0.070 & 0.070 \\
$T_{1u}(3)$ & 0.147 & 0.149 & 0.147 & 0.146 \\
$T_{1u}(4)$ & 0.177 & 0.197 & 0.166 & 0.195 \\
\end{tabular}
\label{table1}
\end{table}

%Figures:

\begin{figure}
\caption{
$f$-strengths per C$_{60}$ from our LDA calculations for the neutral
and charged molecules with up to 6 extra electrons, for the four
T$_{1u}$ vibrational modes. The curves are labeled by the associated
vibrational mode.
}
\label{fig1}
\end{figure}

\begin{figure}
\caption{
Comparison of our calculated strengths with published experimental
measurements
for each of the vibrational modes for (a) neutral C$_{60}$, and
charges
of (b) -1, (c) -3, (d) -4, and (e) -6. The experimental numbers come
from references \protect\onlinecite{Chase} (circles),
\protect\onlinecite{Fu} (bursts),
\protect\onlinecite{Martin} (diamonds), and
\protect\onlinecite{Pichler} (squares).
}
\label{fig2}
\end{figure}

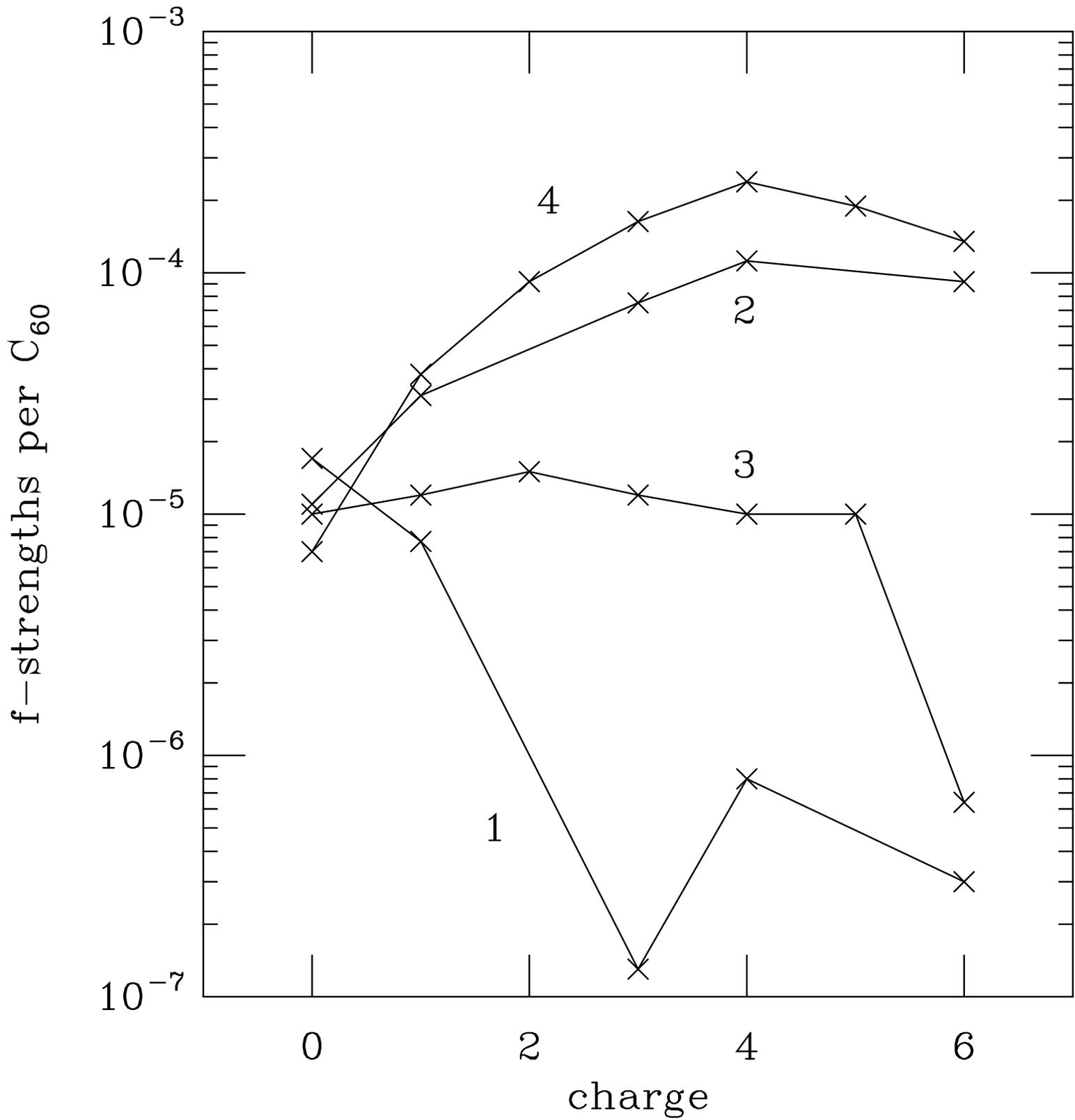
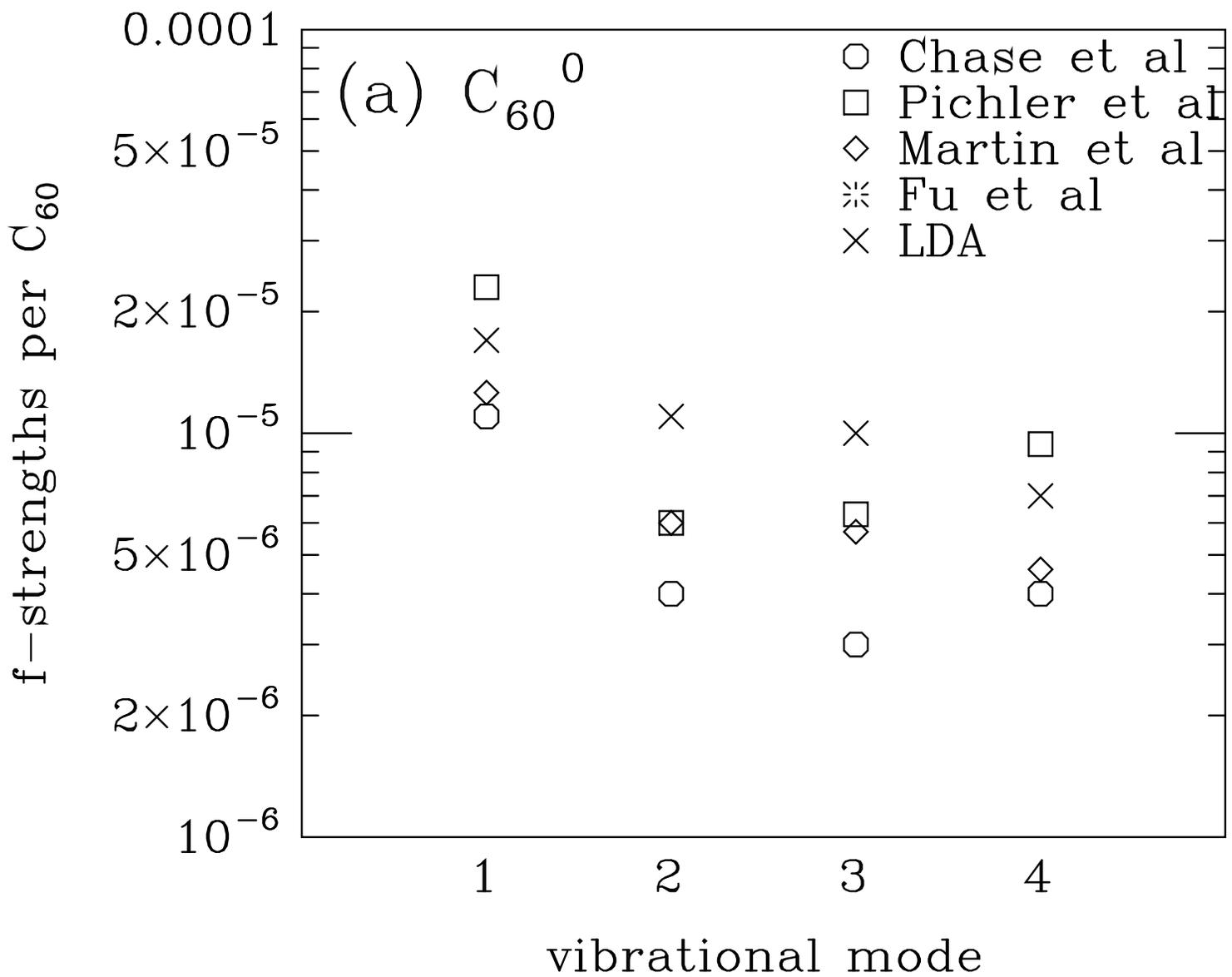
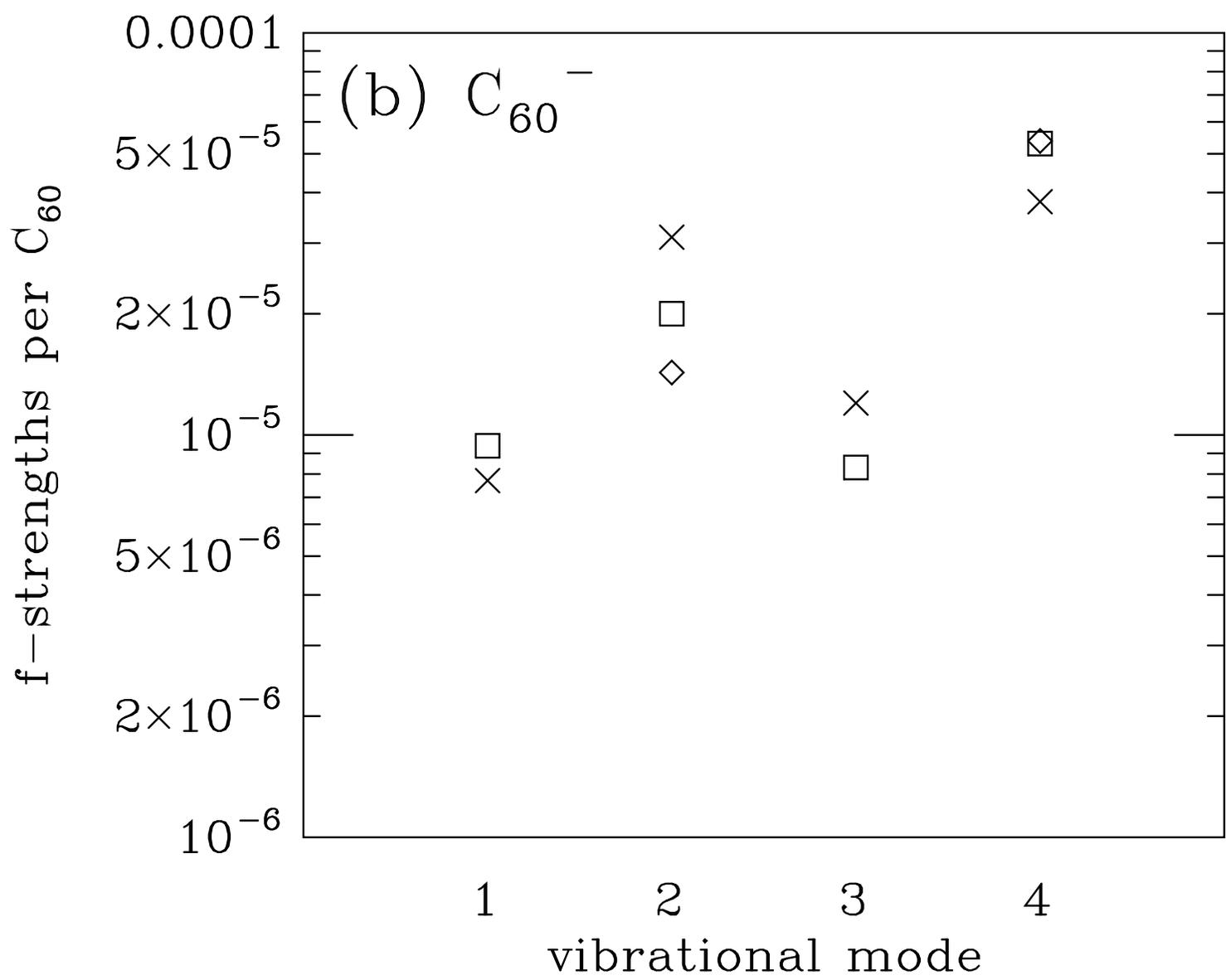
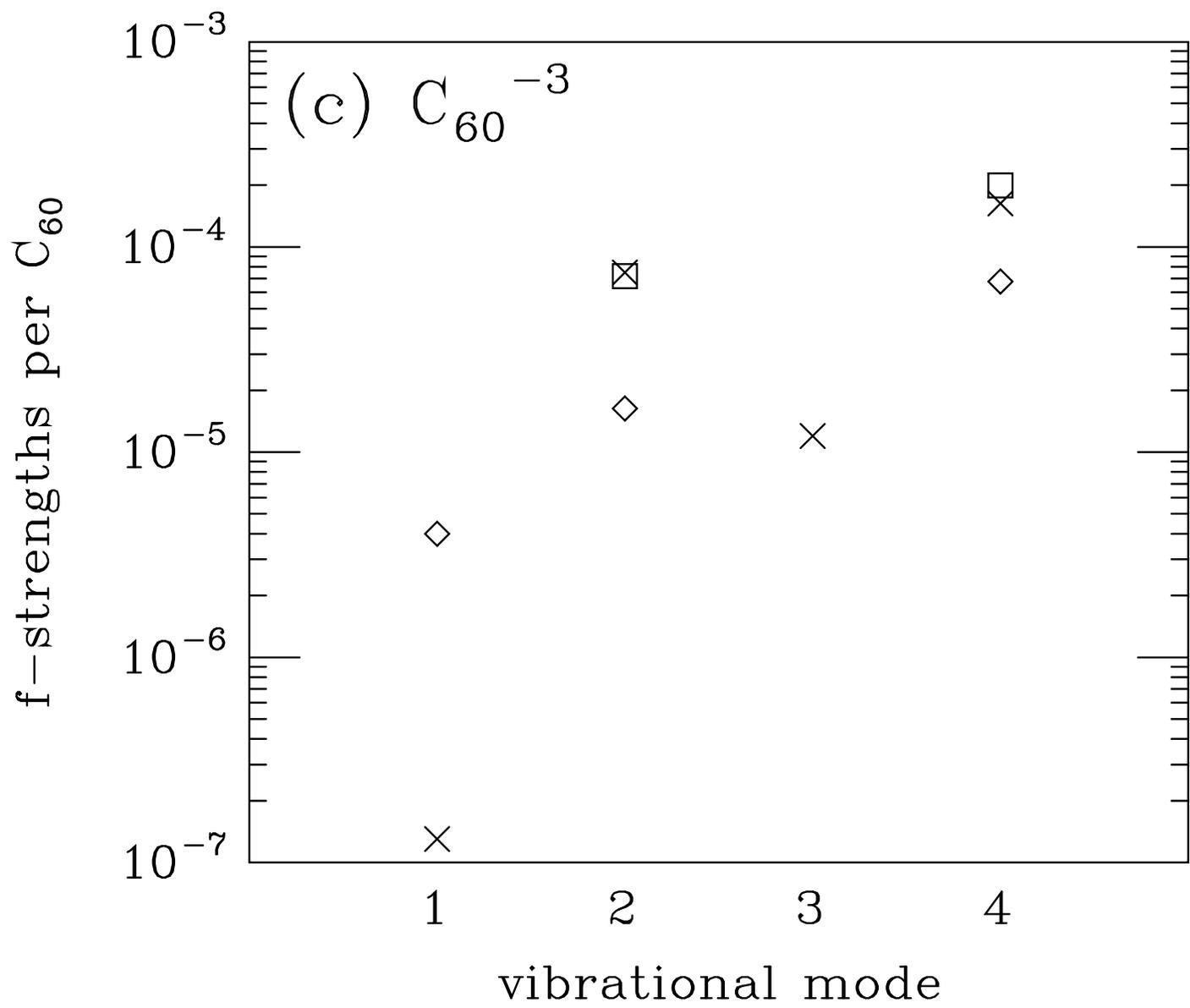
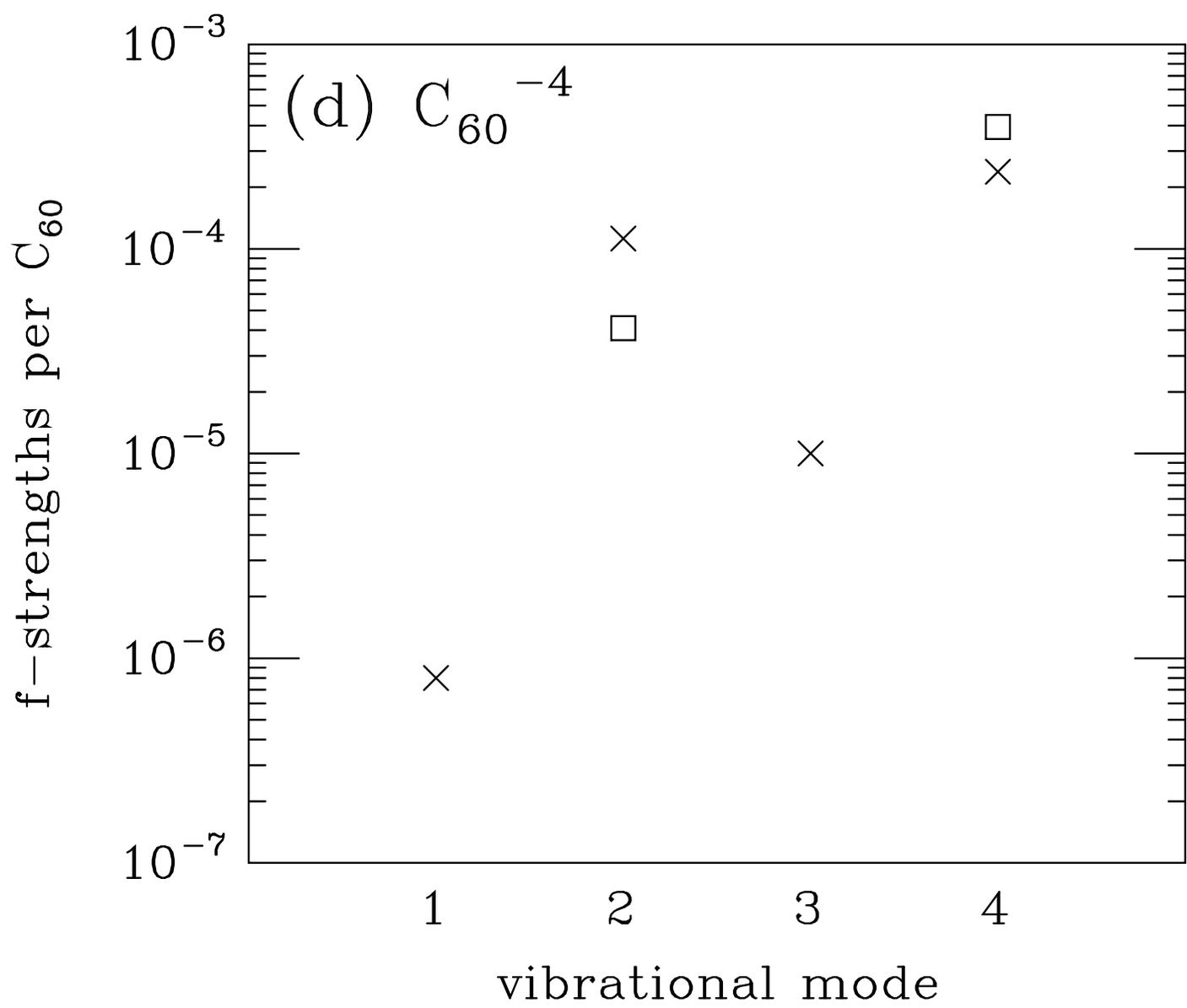
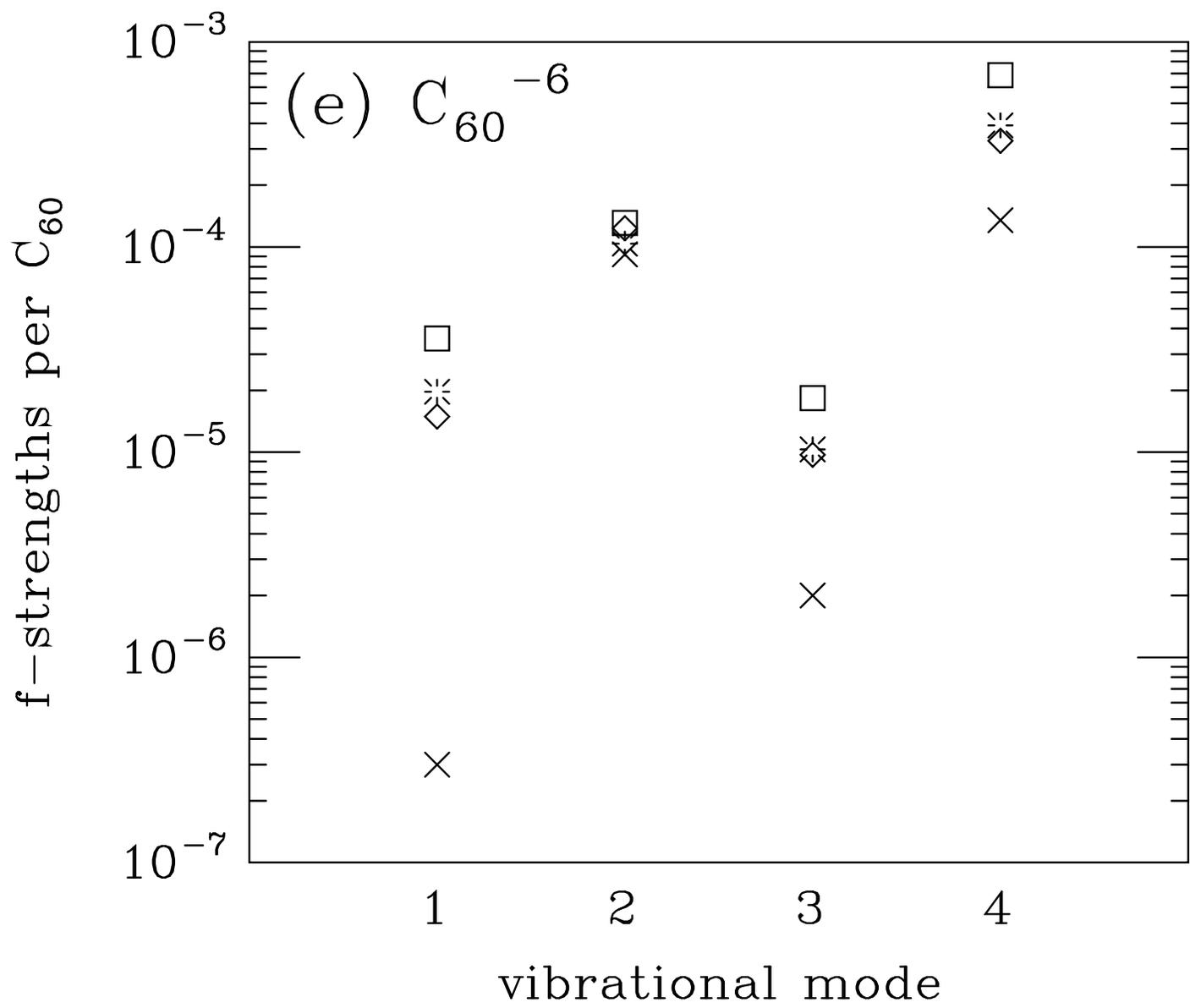
\end{document}